\documentstyle[prd,aps,epsfig]{revtex}
\begin{document}
\setlength{\textwidth}{7.00in}
\setlength{\textheight}{9.0in}
\setlength{\evensidemargin}{-0.2in}
\setlength{\oddsidemargin}{-0.2in}
\setlength{\topmargin}{0.0in}
\input epsf
\draft
\renewcommand{\topfraction}{0.8}
\twocolumn[\hsize\textwidth\columnwidth\hsize\csname
@twocolumnfalse\endcsname \preprint{SU-ITP-00-29, hep-ph/yymmdd}
\title { \Large  \bf  LATTICEEASY\\A Program for Lattice Simulations
of\\Scalar Fields in an Expanding Universe}
\author{Gary Felder$^{1,2}$ and Igor Tkachev$^3$}
\address{${}^1$Department of Physics, Stanford University, Stanford, CA
94305, USA}
\address{${}^2$CITA, University of Toronto, 60 St George Str,
Toronto, ON M5S 3H8, Canada}
\address{${}^3$CERN Theory Division, CH-1211 Geneva 23, Switzerland}
\date{November, 2000}
\vspace{-.1in}
\maketitle
\begin{abstract}
We describe a C++ program that we have written and made available for
calculating the evolution of interacting scalar fields in an expanding
universe. The program is particularly useful for the study of
reheating and thermalization after inflation. The program and its full
documentation are available on the Web at
http://physics.stanford.edu/gfelder/latticeeasy. In this paper we
provide a brief overview of what the program does and what it is
useful for.
\end{abstract}

\pacs{\hskip 1.5cm SU-ITP-00-29 \hskip 3.0cm \ hep-ph/0011159}
\vskip2pc]

\section{Introduction}

Studying the early universe requires describing the evolution of
interacting fields in a dense, high-energy environment. The study of
reheating after inflation and the subsequent thermalization of the
fields produced in this process typically involves non-perturbative
interactions of fields with exponentially large occupations numbers in
states far from thermal equilibrium. Various approximation methods
have been applied to these calculations, including linearized analysis
and the Hartree approximation. These methods fail, however, as soon as
the field fluctuations become large enough that they can no longer be
considered small perturbations. In such a situation linear analysis no
longer makes sense and the Hartree approximation neglects important
rescattering terms. What we have learned in the last several years is
that in many models of inflation preheating can amplify fluctuations
to these large scales within a few oscillations of the inflaton
field. Moreover, such large amplification appears to be a generic
feature, arising via parametric resonance in single-field inflationary
models and tachyonic instabilities in hybrid models.

The only way to fully treat the nonlinear dynamics of these systems is
through lattice simulations. These simulations directly solve the
classical equations of motion for the fields. Although this approach
involves the approximation of neglecting quantum effects, these
effects are exponentially small once preheating begins. So in any
inflationary model in which preheating can occur lattice simulations
provide the most accurate means of studying post-inflationary
dynamics.

Over the past several years we have developed a C++ program for doing
such lattice simulations.  A number of studies have already been
published by us and our collaborators using this program.  We are now
making it available on the World Wide Web in the hopes that it will
stimulate and aid further research in early universe cosmology. We
call the program LATTICEEASY; its website is at
http://physics.stanford.edu/gfelder/latticeeasy/. The website for the
program has documentation, including derivations of all the equations
used in the program. Here we present a short summary of what the
program does and what it can be used for. For more details we invite
you to visit the website.

Section \ref{overview} of this paper gives an overview of what the
program is and how it works. Section \ref{equations} describes the
evolution equations solved by the program as well as the setting of
initial conditions. The references section gives a list of papers
currently published using LATTICEEASY results.

\section{Overview}\label{overview}

Although we designed LATTICEEASY for our studies of reheating after
inflation the program can more generally solve the classical equations
of motion for interacting scalar fields, with or without the effects
of the expansion of the universe.

The study of nonlinear scalar field dynamics has a variety of
applications. We have used it to study parametric resonance
\cite{KT1,KT2,KTgw,KLS,KRT2,KKLT,TKKL,FKLT}, the formation of
gravitational waves \cite{KTgw}, phase transitions and the formation
of topological defects \cite{KRT,KKLT,TKKL,FKLT}, thermalization after
reheating \cite{KRT,FK}, and the formation and evolution of disordered
chiral condensates \cite{FGGKLT}. The program could also be used to
study the classical limit of general nonequilibrium quantum field
theories.

Each particular scalar field potential that the program solves is
encoded in a {\it model file}, which is actually a header file read in
by LATTICEEASY. For example, we have created a model file called
{{\ttfamily twofldm.h}} that contains all the necessary equations for
running the potential
\begin{equation}
V = {1 \over 2} m^2 \phi^2 + {1 \over 2} g^2 \phi^2 \chi^2.
\end{equation}
These equations include the potential itself and its first and second
derivatives, all of which have to be provided in the model file. The
process of creating a model file for a particular potential is
described in the documentation.

Aside from the model file, the only other LATTICEEASY file that the
user needs to modify is {{\ttfamily parameters.h}}, which contains all
the parameters for a given run of the program. These include the
number of grid points, the time step, and a number of other general
variables specific to each run.

One of these variables controls the expansion used in the run. The run
can be set to not include expansion, to use a fixed power-law
expansion, or to self-consistently solve the Friedmann equations using
the fields in the simulation. (This latter option makes sense if the
fields are assumed to be the dominant energy component in the
universe, as is typically the case for simulations of reheating.)

LATTICEEASY has built-in routines for outputting means and variances
of the fields, field spectra, the components of the energy density,
overall energy conservation, histograms of the fields, and
two-dimensional histograms of pairs of fields. If expansion is being
calculated then the scale factor and its derivatives may also be
output. Finally there is an option to call additional output routines
from the model file so that when you create a new model you can design
output specific to it. The file {{\ttfamily parameters.h}} includes a
list of parameters controlling which of these outputs will be
generated as well as specifications such as how often to generate
output, how many bins to use in the histograms, and so on.

The figures below illustrate some of the output produced by
LATTICEEASY. The program's output is in the form of ASCII files and
these plots were generated with Mathematica using those files as
input. The LATTICEEASY website includes a set of Mathematica notebooks
for plotting all the output of the program, but this can easily be
done using any standard plotting software.

\begin{figure}
\leavevmode\epsfxsize=.8\columnwidth \epsfbox{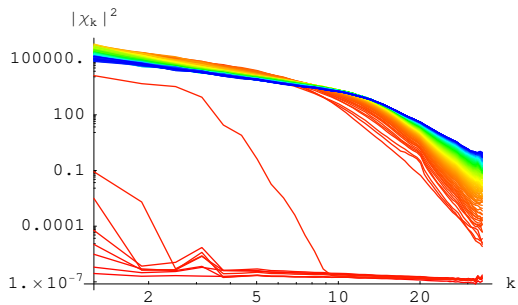}
\caption{The power spectrum of the field $\chi$ in the model $V={1
\over 4} \lambda \phi^4 + {1 \over 2} g^2 \phi^2 \chi^2$. The spectra
rise over time as fluctuations of the field are produced. Initially
the production occurs in resonant peaks but these are quickly smoothed
out.}
\label{evolution}
\end{figure}

\begin{figure}
\leavevmode\epsfxsize=.8\columnwidth \epsfbox{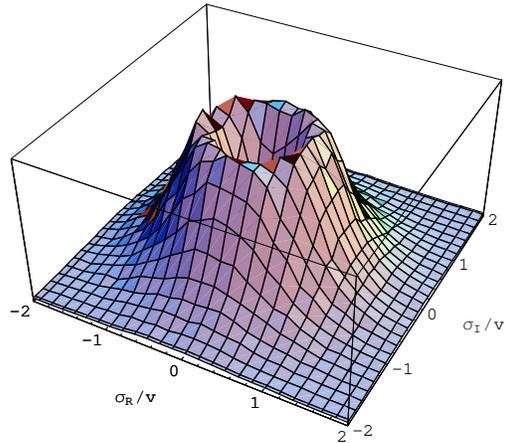}
\caption{Two-dimensional histogram of the fields $\sigma_R$ and
$\sigma_I$ in the SUSY F-Term hybrid inflation model $V = 4 \lambda
\vert\phi\vert^2 \left(\vert\sigma\vert^2 +
\vert\bar{\sigma}\vert^2\right) + 4 \lambda \vert\bar{\sigma} \sigma -
v^2 \vert^2$. The complex field $\sigma$ has fallen down to the
minimum of its potential at $\vert\sigma\vert=v$.}
\label{histogram2}
\end{figure}

Once all of these parameters have been set you simply compile and run
LATTICEEASY. The code is designed to be platform independent and
should work with any C++ compiler.\footnote{It's worth noting as a
warning that the C libraries used by gcc in RedHat Linux 5.0 have a
bug that occasionally causes the program to crash. This is a compiler
bug that was fixed in subsequent versions of gcc, e.g. the
distribution that accompanies RedHat 6.0. Moreover this bug has never
in our experience caused the program to produce incorrect output. When
it occurs it causes the program to crash during the setting of initial
conditions.}

\section{Equations}\label{equations}

LATTICEEASY uses a staggered leapfrog algorithm with a fixed time
step. This means that at each step the field values $f$ and their
derivatives $\dot{f}$ are stored at two different times $t$ and
$t+dt/2$ respectively. The derivatives are used to advance the field
values by a full step $dt$ and then the field values are used to
calculate the second derivatives $\ddot{f}$, which are in turn used to
advance the field derivatives by $dt$. This evolution is done in
place, meaning the newly calculated field values and/or derivatives
overwrite the old ones.

This method is stable for second-order differential equations provided
they have no first derivative terms, i.e. $\ddot{f} = \ddot{f}(f)$. As
we will see below the evolution equations for scalars in an expanding
universe do in general contain first derivative terms, so the program
uses rescaled variables in which these terms are eliminated. These
rescalings are described briefly below and in detail in the
documentation.

In this section we always use $f$ to denote any scalar field and we
use units where $M_p \left((\approx 1.22 \times 10^{19} GeV\right)=1$.

\subsection{Evolution Equations}

The evolution equations solved by LATTICEEASY are simply the
Euler-Lagrange equations for scalar fields in an expanding universe
\begin{equation}
\ddot{f} + 3 H \dot{f} + {\partial V \over \partial f} = 0
\end{equation}
where $f$ is a scalar field and $H = {\dot{a} \over a}$ is the Hubble
parameter. In addition, if expansion is being calculated
self-consistently the program solves the Friedmann equations
\begin{equation}
\ddot{a} = -{4 \pi a \over 3} (\rho + 3 p)
\end{equation}
\begin{equation}
\left({\dot{a} \over a}\right)^2 = {8 \pi \over 3} \rho.
\end{equation}
Either of these Friedmann equations would suffice to solve for the
expansion. (Either Friedmann equation plus the field evolution
equations implies the other one.) In practice LATTICEEASY uses a
combination of the two chosen for computational convenience.

The field and spacetime variables used by the program are rescaled
from their bare, physical values. These rescalings accomplish several
things: They eliminate the $\dot{f}$ term from the field evolution
equations, thus making the staggered leapfrog method stable. They can
simplify the equations by eliminating coupling constants or other
parameters. They can set the field and time values to the natural
scales of the problem, e.g. by having the time variable correspond to
the number of inflaton oscillations. The program has algorithms for
determining the most convenient rescalings for a given potential. If
you wish, however, you can set the rescalings manually to whatever
scales you find most convenient. The program does impose, however, a
certain relationship between the field and spacetime rescalings that
ensures that $\dot{f}$ will be eliminated from the equations of
motion. See the documentation for details.

\subsection{Initial Conditions}

Although the field equations are solved in configuration space with
each lattice point representing a position in space, the initial
conditions are set in momentum space and then Fourier transformed to
give the initial values of the fields and their derivatives at each
grid point. The initial field values are given by quantum fluctuations
with a dispersion characterized by
\begin{equation}
<\vert f_k\vert^2> = {1 \over 2 \omega_k}.
\end{equation}
where $f_k$ is the Fourier transform of $f$ and
\begin{equation}
\omega_k^2 = k^2 + m^2,
\end{equation}
\begin{equation}
m^2 = {\partial^2 V \over \partial f^2}.
\end{equation}
The phase of each mode is set randomly and the amplitude of each mode
comes from a Gaussian random distribution. One of the run parameters
is a random number seed that can be adjusted to do runs with different
randomized initial conditions.

In addition to these fluctuations the program allows you to set
homogeneous initial values for the fields and derivatives. The initial
field value at each point is a sum of the homogeneous value and the
fluctuations at that point.

\section{Conclusions}

There are several additions to the program that we hope to make in the
future. A great deal of attention has been paid recently to the growth
of metric perturbations in the early universe, and we would like to
extend LATTICEEASY to calculate the coupled equations for the growth
of field and metric fluctuations. We also anticipate adding vector
fields to the lattice calculations, thus allowing the program to
calculate gauge models. We have thought about incorporating fermionic
fields as well, but this presents some difficulties because fermionic
fields have no well-defined classical limit. At the moment it's
uncertain if we will find a way to include them in our calculations or
not.

The list of references below shows the work that has already been
published using the results of LATTICEEASY. As computing power
increases and cosmology turns to increasingly difficult mathematical
problems, the role of large scale numerical simulations in the field
is likely to keep increasing dramatically. Given our uncertainty about
the correct particle physics models to use for describing the early
universe, we believe that the most important tools for numerical
computation will be ones that are flexible enough to easily accomodate
a wide variety of models. We designed LATTICEEASY to have this
flexibility while still remaining easy to use and understand.

\section*{Acknowledgements}

We would like to thank Sergei Khlebnikov, Lev Kofman, and Andrei Linde
for valuable discussions and contributions to this
project. G.F. thanks CITA for its hospitality during much of the time
the program was written.

\end{document}